\documentclass[a4paper,10pt]{article}
\usepackage{amstex}
\usepackage{epsfig}

\makeatletter
\renewcommand{\section}{\@startsection{section}{1}{0em}{-\baselineskip}%
{\baselineskip}{\normalfont\large\bfseries}}
\renewcommand{\subsection}%
{\@startsection{subsection}{2}{0em}{-0.7\baselineskip}%
{0.7\baselineskip}{\normalfont\bfseries}}
\makeatother

\newcommand{\GeV}{\,\mathrm{GeV}}

\newcommand{\ecm}{e\,\mathrm{cm}}

\newcommand{\M}[1]{M_{\mathrm{#1}}}
\newcommand{\J}[1]{J_{\mathrm{#1}}}

\newcommand{\im}{\operatorname{Im}}
\newcommand{\tr}{\operatorname{Tr}}
\newcommand{\lau}
{\boldsymbol{\lambda}_{\text{\textsc{u}}}^{\phantom{\dagger}}}
\newcommand{\lad}
{\boldsymbol{\lambda}_{\text{\textsc{d}}}^{\phantom{\dagger}}}
\newcommand{\lae}
{\boldsymbol{\lambda}_{\text{\textsc{e}}}^{\phantom{\dagger}}}
\newcommand{\laun}
{\boldsymbol{\lambda}_{\text{\textsc{u}}}}
\newcommand{\ladn}
{\boldsymbol{\lambda}_{\text{\textsc{d}}}}

\newcommand{\laud}
{\boldsymbol{\lambda}_{\text{\textsc{u}}}^\dagger}
\newcommand{\ladd}
{\boldsymbol{\lambda}_{\text{\textsc{d}}}^\dagger}
\newcommand{\laed}
{\boldsymbol{\lambda}_{\text{\textsc{e}}}^\dagger}
\newcommand{\ped}[1]{{\mathnormal{#1}}}
\newcommand{\glu}{{\tilde{g}}}
\newcommand{\glum}{3}
\newcommand{\mybf}[1]{\boldsymbol{#1}}
\newcommand{\ord}[1]{\mathcal{O}(#1)}

\newcommand{\C}{^\dagger}
\newcommand{\NP}[3]{{Nucl.\ Phys.\ \bf  #1}, #3 (#2)}
\newcommand{\PRL}[3]{{Phys.\ Rev.\ Lett. \bf #1}, #3 (#2)}
\newcommand{\PL}[3]{{Phys.\ Lett.\ \bf #1}, #3 (#2)}
\newcommand{\PR}[3]{{Phys.\ Rev.\ \bf #1}, #3 (#2)}

\newcommand{\Frac}[2]{\leavevmode\kern.1em\raise.5ex
\hbox{\the\scriptfont0 #1}
\kern-.1em/\kern-.15em\lower.25ex\hbox{\the\scriptfont0 #2}}
\def\diag{\mathop{\rm diag}}
\def\circa#1{\,\raise.3ex\hbox{$#1$\kern-.75em\lower1ex\hbox{$\sim$}}\,}

\newcounter{mysubequation}[equation]
\renewcommand{\themysubequation}{\alph{mysubequation}}
\newcommand{\mytag}%
{\stepcounter{mysubequation}\tag{\theequation\themysubequation}}
\newcommand{\restore}{\stepcounter{equation}}

\newcommand{\preprintdate}{October 1996}
\newcommand{\preprintnumber}{IFUP--TH 63/96}
\newcommand{\hepnumber}{arch-ive/9610485}
\newcommand{\titletext}{
{\normalsize\emph{October 1996}      \hfill \bfseries{IFUP--TH 63/96}}\\[-1ex]
{\normalsize\bfseries{hep-ph/9610485} \hfill  FT--UAM 96/25}\\[2ex]
Electric dipole moments from Yukawa
phases \\ in supersymmetric theories}  
\newcommand{\abstracttext}{We study quark and electron EDMs generated
by Yukawa couplings in
supersymmetric models with different gauge
groups, using the EDM properties under flavour
transformations. In the
MSSM (or if soft terms are mediated below the unification scale)
the one loop contributions to the neutron EDM are smaller than in
previous computations based on numerical methods, although increasing
as $\tan^3\!\beta$. 
A neutron EDM close to the experimental limits can be generated in SU(5), if
$\tan\beta$ is large, through the $u$-quark EDM $d_u$, 
proportional to $\tan^4\!\beta$.
This effect
has to be taken into 
account also in SO(10) with large $\tan\beta$, where $d_u$ is
comparable to the $d$ quark EDM, proportional to $\tan\beta$.
}
\newcommand{\authortext}{\bigskip
\textbf{Andrea Romanino$^{\, a}$ \textmd{and} Alessandro
Strumia$^{\, b}$}
\medskip\\
\em\normalsize $\mbox{}^a$ Dipartimento di Fisica, Universit\`a di
Pisa and INFN, \\Sezione di Pisa, I-56126 Pisa,
Italia\\[0.1\baselineskip] 
$\mbox{}^b$ Departamento de F\'{\i}sica Te\'orica,
Universidad Aut\'onoma de Madrid, 28049,
Madrid, Espa\~na and INFN, Sezione di Pisa, I-56126 Pisa, Italia
}

\newcommand{\makeonlyfirstpage}{%
\setlength{\topmargin}{21mm}
\setlength{\textheight}{172mm}
\renewcommand{\footnoterule}{} 
\title{%
\normalsize\hspace*{\fill}
\begin{tabular}{l}\preprintnumber\\\hepnumber\end{tabular}
\vspace{3\baselineskip}\\\huge\bfseries\titletext}
\author{
\begin{minipage}[t]{0.8\textwidth}
\large\centering\authortext
\end{minipage}}
\date{\preprintdate}
\begin{document}
\maketitle
\begin{abstract}\large\noindent\abstracttext\end{abstract}
\thispagestyle{empty}
\end{document}}

\title{\LARGE\bfseries\titletext}
\author{\begin{minipage}[t]{0.8\textwidth}
\large\centering\authortext
\end{minipage}}
\date{}

\begin{document}

\bigskip
\maketitle
\begin{abstract}\normalsize\noindent\abstracttext\end{abstract}
\normalsize\vspace{\baselineskip}

\section{Introduction}
\label{sec:int}

In supersymmetric theories there are various possibilities of
generating an intrinsic electric dipole moment (EDM) in
microscopic systems.

First of all, the neutron has already a
non vanishing EDM in the Standard Model (SM), due to the
uneliminable phase in the CKM matrix, or, in other words, to the
misalignment between the left eigenstates of the two Yukawa matrices,
$\laun$ and $\ladn$, and to the charged current gauge interactions
between the left-handed quarks, that make the misalignment physically
significant. 

If further interactions are present, further misalignments and phases
can become significant.
For example, in the unified extensions of the SM the new gauge
interactions make significant the misalignments
among particles unified in the same gauge multiplet.
Nevertheless, without supersymmetry, due to the decoupling of heavy
particles, these effects are suppressed at low energy by powers of the
unification mass.

In the supersymmetric extensions of the SM, one more possible source of
flavour and CP violation is associated to the possibility that the
scalar partners of quarks and leptons are not degenerate and
point in 
different directions in the flavour space compared to the
corresponding fermions.
On the other hand, models are often considered in which the soft
supersymmetry 
breaking terms are flavour universal and real at some energy scale $M_0$.
This can happen in models in which supersymmetry breaking is communicated
to the observable sector by 
gravity~\cite{barbieri:82a,moduli} 
(in this case $M_0=\M{P}\equiv\text{reduced Planck mass}$)
or when it is communicated by gauge interactions at relatively low
energy~\cite{alvarez:82a}. These 
hypotheses allow to keep under control flavour changing neutral
current processes and EDMs due to the mixings in the
fermion-scalar interactions.
In this case, effects coming from the non universal radiative
corrections to soft breaking parameters
have to be ascribed to the Yukawa couplings.

As said, when the gauge group is unified, some new phases and flavour
mixings otherwise inexistent or unphysical can become significant.
Unlike what happens without supersymmetry, if its 
breaking is communicated to the observable sector at a scale where the
gauge group is already unified, these further phases and flavour
mixings can succed in contaminating the soft breaking matrices before
the heavy particle decoupling make them harmless, giving rise to
important effects~\cite{barbieri:94a,dimopoulos:95a}.

Besides the quark and lepton sector and their scalar partner
sector, a CP source can come from the hidden supersymmetry breaking
sector of the theory. As 
a consequence, there can be CP violating effects in the observable
sector not associated to a mixing matrix and to a consequent flavour
violation.
As a matter of fact, even in the flavour universality hypothesis, a CP
violation in the hidden sector can make the soft breaking parameters
complex since the beginning. In this case, the effects are very large
unless the phases are forced to be small.

In this paper, we consider the effects of
phases associated to the misalignment of two Yukawa matrices, $\laun$
and $\ladn$, in the
hypothesis that the soft supersymmetry breaking parameters are
generation universal and real at an high energy scale $M_0$.
We assume that the Yukawa interactions are present until $M_0$ without
worring about their origin. 
However the minimal unified models that we consider
do not give correct unification predictions for the
light fermion masses, showing that, differently from the MSSM case,
flavour physics cannot be totally decoupled from unification physics.
Furthermore, minimal flavour
effects could be comparable to minimal unification
effects~\cite{goldberg:96a}.  

The order of magnitude of EDMs for the quarks $u$, $d$ and for the
electron $e$ depends in a crucial way on the
gauge structure at the universality scale $M_0$. We will consider
therefore different gauge structures, from the minimal one
(SU(3)$\times$SU(2)$\times$U(1)) to that one that unifies all fermions
belonging to the same family (SO(10)), passing through the minimal
unification (SU(5)), studying them from the point of view of the
properties of EDMs under flavour transformations 
commuting with the gauge ones.
This will allow us to give our estimates for quark and electron
EDMs.

We will consider the MSSM in section~\ref{sec:MSSM}.
In Refs.~\cite{bertolini:94a} 
the calculation of the down quark EDM induced by a loop with charginos
has been made using numerical methods with sometimes different results.
For a given value of $\tan\beta$, $\tan\beta=10$, Bertolini and
Vissani claim in fact that $d_n$ can be as large as
$d_n=\ord{10^{-30}\ecm}$, 
Inui et al.\ obtain 
$d_n=\ord{10^{-29}\mbox{--}10^{-27}\ecm}$ and Abel et
al.\ get $d_n=\ord{10^{-33}\mbox{--}10^{-29}\ecm}$,
all of them indicating a linear dependence on $\tan\beta$. The
numerical calculation is used in order to solve the matricial
renormalization group equations for the Yukawa couplings and the soft
breaking terms; this is necessary for doing a detailed analysis at the
Fermi scale.
Unfortunatly, in the low energy lagrangian, the particular flavour
structure of the CKM CP violation, at the basis of the smallness of
the EDMs, is hidden in a large number of flavour mixing matrices.
On the contrary, by studying the EDMs directly in terms of the parameters
at the universality scale, it is easy: i) to estimate the order of
magnitude of the effect, that for $\tan\beta=10$ does not much exceed
$10^{-33}\ecm$, ii) to display the cancellation of contributions not
containing the squared light Yukawa couplings, that suppress the EDMs,
iii) to show the raise with $\tan^3\!\beta$ of the $d$ quark EDM.
It is also possible to obtain the relation $d_d/m_d=-d_s/m_s$ between
the one loop dipoles of quarks $d$ and $s$, $d_d$ and $d_s$ (that both
contribute to $d_n$). 

From our estimates we see that the one loop supersymmetric effects
are comparable with the SM ones~\cite{dNreview,dNexch}.  It turns out
that also supersymmetric contributions at higher loops are
relevant.

In section~\ref{sec:unified} we will consider unified theories.
The possibility of EDMs for the neutron and the electron close to
experimental limits in unified theories has been pointed out and
analyzed in Refs.~\cite{dimopoulos:95a,barbieri:95a} in
the case of moderate $\tan\beta$, where it is associated to the gauge
group SO(10). The big 
enhancement of the neutron EDM in comparison with the MSSM case is
due to the fact that, unlike in the MSSM,  in a unified theory CP can
be violated also when 
two eigenvalues of the same Yukawa matrix are
vanishing~\cite{romanino:96a}. 
Nevertheless this, as other effects~\cite{arkani:96a}, happens also in
the SU(5) case, but only when the $u$ quark is concerned. 
As the third generation Yukawa coupling involved is the bottom one,
the effect is proportional to $\tan^4\!\beta$. 
Due to its possible interest, we also present the results of a numeric
computation of $d_u$.
The $d$ quark contribution, as the electron EDM, turns out to be very
small compared with $d_u$. We conclude section~\ref{sec:unified} with
a short rivisitation of SO(10).
In appendix, we give the general formulas for the one loop EDMs of up
and down quarks and charged leptons
in terms of the parameters of the Lagrangian at the Fermi scale.

\section{EDMs in the MSSM}\label{sec:MSSM}
In this section we will study quark and electron EDMs  in the minimal
supersymmetric extension of the Standard Model, with 
real and generation-universal soft terms at some scale $M_0$
\[
\begin{array}{cc}
\begin{gathered}
  \mybf{A}_\ped{f} = A_{\ped{f},0}\boldsymbol{1}, 
  \quad \ped{f}=\ped{u},\ped{d},\ped{e}\\ 
  \mybf{m}^2_{\tilde{\ped{R}}} =
  m^2_{\tilde{\ped{R}},0}\boldsymbol{1}  
  \quad \ped{R}=\ped{Q},\ped{u}_R,\ped{d}_R,\ped{L},\ped{e}_R\\
\end{gathered} & \text{at $M_0$.}
\end{array}
\]
This model provides the minimal amount of CP violation and,
consequently, the
minimal contribution to $d_n$  in a
supersymmetric theory with soft breaking terms generated at $M_0$.  

\medskip

In the described hypothesis, third generation Yukawa couplings cannot
be, by themselves, a source of quark EDMs. 
In fact, let us suppose that we neglect the four light Yukawa couplings
in the quark sector, $\lambda_u,\lambda_c,\lambda_d,\lambda_s$
everywhere, a part for the 
mass that must be present in a loop diagram in order to provide the
elicity flip. Then, since the hypothesis is scale independent, CP
violation disappears and the EDMs have to vanish.
In fact, at $\M{P}$ the soft terms are universal and real,
so that CP violation can only come from Yukawa couplings. 
But, as there
is a couple of degenerate eigenvalues in each Yukawa matrix, all
CKM phases can be eliminated using rotations in the 1-2 sector and
phase redefinitions of the fields, just as in the
Standard Model.
More precisely, in the $d_u$ case we have to use a
rotation of {\em down} left light  generations, while in the $d_d$
case we have to use a rotation of {\em up} ones. This is 
because when we consider, for example, the EDM of the quark $d$,
the $d_L$ mass eigenstate is fixed, and the possibility of rotating
the $d_L$ and $s_L$ eigenstates is lost.
In the limit of vanishing $\lambda_d$ and $\lambda_s$, CP violation
disappears from the theory, but also the lighest down mass eigenstate
becomes not defined.

Thus the light Yukawa couplings play a crucial role in
generating the 
EDMs. Let us now see in which way they intervene. 
To begin with a simpler case, let us consider the imaginary part of
the $B$ term (defined in appendix) after one
loop rescaling from $M_0$ to $M_Z$. This quantity, like the EDMs,
vanishes in 
absence of CP violation and contributes to the EDMs themselves.
The $B$ term does not depend on the basis in the flavour space in
which the left doublets $\ped{Q}$, $\ped{L}$ and the right singlets
$\ped{u}^c$, $\ped{d}^c$, 
$\ped{e}^c$ are written.
In other words, if we consider a transformation of the flavour
components of the superfields $\hat{\ped{Q}}$, $\hat{\ped{u}}^c$, 
$\hat{\ped{d}}^c$, $\hat{\ped{L}}$, $\hat{\ped{e}}^c$ 
commuting with the gauge group, namely a
$\text{U(3)}^5 = \text{U(3)}_\ped{Q}\times \text{U(3)}_{\ped{u}^c}
\times \text{U(3)}_{\ped{d}^c}\times \text{U(3)}_\ped{L}\times
\text{U(3)}_{\ped{e}^c}$ transformation, the Yukawa couplings become
\begin{equation}
\label{trans}
\lau\rightarrow U^T_{\ped{u}^c}\lau U_\ped{Q},\qquad
\lad\rightarrow U^T_{\ped{d}^c}\lad U_\ped{Q},\qquad
\lae\rightarrow U^T_{\ped{e}^c}\lae U_\ped{L},
\end{equation}
whereas the $B$ term remains invariant. For what follows,
most important are 
the $\text{U(3)}_\ped{Q}\times \text{U(3)}_{\ped{u}^c}\times
\text{U(3)}_{\ped{d}^c}$  transformations. Since at $M_0$ all
the parameters in the tree level lagrangian except the Yukawa couplings
are invariant, it is convenient to consider the $B$ term at the
Fermi scale as a function of those parameters. Actually, the relation
between high and low energy Yukawa couplings can be inverted, so that
the Yukawa couplings can be considered at the Fermi scale.

Owing to the invariance relative to transformations of the
right-handed quarks, $B$ depends on the Yukawa couplings only through
their squares 
$\laud\lau$, $\ladd\lad$ and $\laed\lae$. Moreover, if two among the
$\lau$ or $\lad$ eigenvalues are equal at $M_0$, the Yukawa
couplings can be made real through a transformation~\eqref{trans} so
that $\im B$ vanishes. Therefore the RGE corrections to $\im B$ must
be proportional
to $(\lambda_t^2-\lambda_c^2) (\lambda_t^2-\lambda_u^2)
(\lambda_c^2-\lambda_u^2) (\lambda_b^2-\lambda_s^2)
(\lambda_b^2-\lambda_d^2) (\lambda_s^2-\lambda_d^2)
\approx \lambda_t^4 \lambda_b^4 \lambda_c^2 \lambda_s^2$.
In fact, owing to the invariance relative to a generic $U(3)^5$
transformation, $B$ is a sum of terms like
\begin{equation}
\tr\bigl[(\laud\lau)^{n_1}(\ladd\lad)^{m_1}\dotsm
(\laud\lau)^{n_r}(\ladd\lad)^{m_r}\bigr]
\tr\bigl[(\laed\lae)^k\bigr]
A_{u,d}^0
\end{equation}
with real adimensional functions of $M_0$ and $A_{u,d}^0$ 
as coefficients. The first non vanishing supersymmetric contribution
to $\im B$ is then proportional to
\begin{equation}\label{imb}
\tr\bigl[(\laud\lau)^2 (\ladd\lad)^2
(\laud\lau)(\ladd\lad)  \bigr] (A_u^0-A_d^0) 
\simeq \lambda_t^4 \lambda_b^4\lambda_c^2 \lambda_s^2 
\J{CP}(A_u^0-A_d^0) ,
\end{equation}
where $\J{CP}=\im (V^{\dagger}_{dt} V^{}_{tb}
V^{\dagger}_{bc} V^{}_{cd})$. 
If $A_u^0=A_d^0$, $\im B$ is furtherly suppressed by lepton Yukawa
couplings or by small effects due to the $u_R$-$d_R$ hypercharge
difference.

\medskip

Let us consider now the imaginary part of a flavour non-invariant quantity,
more precisely the imaginary part of the matrix element of a quantity
$D$ 
transforming as $\lad$ (or of a quantity $U$ transforming as $\lau$),
calculated in correspondence of left and right mass eigenstates.
The imaginary parts appearing in the expressions~\eqref{expr} in
appendix for the
EDMs and thus the EDM themselves are examples of such quantities.
In this case, the light Yukawa coupling suppression is less strong and
dependent on the mass eigenstate we consider, up or down, light or
heavy.

The general dependence of $D$ on the Yukawa couplings is
$D = \lad \cdot f(\laud\lau,\ladd\lad,\tr)$, where $\tr$
represents traces of $(\laed\lae)^k$ and we suppose that $f$
is a real polynomial in the Yukawa couplings, as it is
for the one loop EDMs, where the Yukawa couplings
only come from vertices or supersymmetric RGE corrections. The
imaginary part of the $d_R$-$d_L$ matrix element is then proportional
to the $d$ eigenvalue
\begin{equation}
\label{DdR}
\im\bigr[D_{d_R d_L}\bigl] = \lambda_d 
\im\bigl[ f\bigl(\laud\lau,\ladd\lad, \tr)_{d_L d_L}\bigr].
\end{equation}
Moreover, in getting the dependence of $f(\ldots)_{d_L d_L}$ on the
light Yukawa couplings, it is no longer possible to consider the limit
$\lambda^2_d=\lambda^2_s$ or $\lambda^2_d=\lambda^2_b$ as above for
$B$, because in this limit the eigenstate $d_L$, as $d_R$, is no
longer defined.
Therefore $f(\ldots)_{d_L d_L}$ has not to be proportional to
$\lambda^4_b\lambda^2_s$ but only to $\lambda^2_b$, so that the
necessary dependence on the Yukawa couplings is $\im\bigr[D_{d_R
d_L}\bigl] \propto \lambda^4_t\lambda^2_c\lambda^2_b\J{CP}$.
The first non vanishing contribution to $\im\bigr[D_{d_R d_L}\bigl]$
comes in fact from the term proportional to $\lad
(\laud\lau)^2
(\ladd\lad)
(\laud\lau)$ in the
expression for $D$. Let $a$ be the proportionality coefficient.
Since higher order terms are negligible, we have
\begin{equation}\label{omin}
\im\bigl[D_{d^R_i d^L_i}\bigr]\simeq
a\lambda_{d_i}\im\bigl[
(\laud\lau)^2
(\ladd\lad)
(\laud\lau)
\bigr]_{d^L_i d^L_i}.
\end{equation}
Had we considered a matrix $U$ transforming as $\lau$, we would have
found
\begin{equation}\label{omin2}
\im\bigl[U_{u^R_i u^L_i}\bigr]\simeq
b\lambda_{u_i}\im\bigl[
(\ladd\lad)^2
(\laud\lau)
(\ladd\lad)
\bigr]_{u^L_i u^L_i}.
\end{equation}
Since the flavour dependence of the one loop EDMs is all contained in
the 
arguments of imaginary parts in~\eqref{expr}, all the down EDMs are
expressable in the previous form $d_{d_i}=\im\bigl[D_{d^R_i
d^L_i}\bigr]$ with the same $D$, so as for the up ones we have
$d_{u_i}=\im\bigl[U_{u^R_i u^L_i}\bigr]$ 
with the same $U$.
Then there are precise and parameter
independent relations between one-loop
supersymmetric contributions to EDMs 
of quarks of different families\footnote{We remind that
$\im(V\C b V c V\C d V)_{11} =
c_{32}[b_{21}d_{31}-b_{31}d_{21}]J_{\rm CP}$ 
where $b,c,d$ are diagonal flavour matrices,
$b=\diag(b_1,b_2,b_3)$, $b_{ij}\equiv b_i-b_j$, etc.}:
\begin{alignat}{2}
\label{ddD}
d_d &\simeq +a\lambda_d\lambda_t^4\lambda_c^2\lambda_b^2\J{CP} &\qquad
d_u &\simeq -b\lambda_u\lambda_b^4\lambda_s^2\lambda_t^2\J{CP}  
\mytag\\
d_s &\simeq -a\lambda_s\lambda_t^4\lambda_c^2\lambda_b^2\J{CP} &\qquad
d_c &\simeq +b\lambda_c\lambda_b^4\lambda_s^2\lambda_t^2\J{CP}  
\mytag\\
d_b &\simeq +a\lambda_b\lambda_t^4\lambda_c^2\lambda_s^2\J{CP} &\qquad
d_t &\simeq -b\lambda_t\lambda_b^4\lambda_s^2\lambda_c^2\J{CP}  
\mytag
\end{alignat}
\restore
so that
\begin{align}
\label{dsm}
d_s&=-\frac{m_s}{m_d}d_d=\frac{m_b}{m_s}d_b\mytag\\
d_c&=-\frac{m_c}{m_u}d_u=\frac{m_t}{m_c}d_t.\mytag
\end{align}
\restore
Eqs.~\eqref{dsm} show that 
the middle generations EDMs are the largest ones and 
allow to express the neutron EDM in terms of only the $u$ and $d$
quark 
EDMs. Using eqs.~\eqref{dn1} and \eqref{dsm}, and neglecting
chromoelectric dipole moment contributions, we get in fact the
estimate
\begin{equation}
\label{dn2}
d_n = x d_d +y d_u,\quad\mbox{with}\quad x\approx 4.6\mbox{ and } 
y\approx 0.7.
\end{equation}

From eqs.~\eqref{ddD} it is apparent that
$d_d/m_d+d_s/m_s+d_b/m_b$ is much smaller than $d_d/m_d$. This is
because the previous combination is invariant under flavour rotations of
the quark fields and thus suppressed by both $\lambda^2_c$ and
$\lambda^2_s$, just as the 
RGE induced phase of $B$, electron EDM, strong CP angle and 3-gluon
operator~\cite{WeinbergGGG}.
Owing to this large suppression, the $B$ term phase contributes to
the neutron EDM in a negligible way.

Eqs.~\eqref{ddD} show the strong rise of the EDMs with
$\tan\beta$. The large $\tan\beta$ region is therefore by far the most
interesting one and it is the one that we will consider in the
following. Expressing all fermion masses in terms of the Yukawa
couplings, 
the weak vacuum expectation value $v=174\GeV$ and $\tan\beta$, the only
dependence of EDMs on $\tan\beta$ is in the Left-Right (and R-L) blocks of the
scalar mass matrices that provide the ``elicity flip'' in the scalar
sector%
.
In the large $\tan\beta$ region,
\begin{equation}
{M^2_D}_{RL}\simeq-v\mu\lad,\qquad
{M^2_U}_{RL}\simeq-v\lau\mybf{A}_\ped{u}
\label{RL}
\end{equation}
so that the EDMs depend on 
$\tan\beta$ only through the down Yukawa couplings: 
$\lad\propto 1/\cos\beta\simeq\tan\beta$.
Hence the down quark EDMs increase with $\tan^3\!\beta$, whereas the up
quark ones increase with $\tan^6\!\beta$.

Let us consider now more closely how one loop graphs can give rise to
the flavour structure of eqs.~\eqref{ddD}.
In both the $d_d$ and $d_u$ cases there are two relevant graphs (see
appendix) involving either charginos or gluinos.
The Yukawa matrices necessary to obtain the flavour structure of
eqs. \eqref{omin} and \eqref{omin2} come from the RGE corrections to
the soft breaking parameters and (in the case of chargino exchange
only) from the vertices.
Each insertion of a couple of Yukawa coming from RGEs is accompanied
by a loop factor and a large logarithm,
$t_Z=(4\pi)^{-2}\ln(M^2_0/M^2_Z)$. In order to provide the 9 Yukawa
couplings of eqs.~\eqref{ddD}, 4 of these inserctions are necessary in
the gluino diagram, giving a factor $t_Z^4$, whereas 3 are enough in
the chargino diagram ($t^3_Z$).

To estimate the one loop contributions to $d_d$ and $d_u$ only the
dependence on dimensionful parameters is missing.
Because of the necessary presence of a L-R scalar mass inserction as
in eq.~\eqref{RL} and of the behavior of the loop functions, it is
\begin{align}\label{dnt}
d_d^{\tilde{h}}&\approx  \frac{e}{(4\pi)^2}t_Z^3
\lambda_d 
\lambda_t^4\lambda_b^2 \lambda_c^2
\frac{v\mu A_u}{\max(m^2_{\tilde{h}}, m^2_{\tilde{u}})m^2_{\tilde{u}}}
J_{\rm CP}\approx 
10^{-31}\:\ecm\:\fracwithdelims(){\tan\beta}{55}^3
\fracwithdelims(){200\GeV}{M_{\rm SUSY}}^2
\mytag\\
d_d^{\tilde{g}}&\approx e\frac{\alpha_3}{4\pi}
t_Z^4\lambda_d
\lambda_t^4\lambda^2_b\lambda_c^2 
\frac{v\mu M_\glum}{\max(M_\glum^2,m_{\tilde{d}}^2)m_{\tilde{d}}^2}
J_{\rm CP}\approx
10^{-32}\:\ecm\:\fracwithdelims(){\tan\beta}{55}^3
\fracwithdelims(){200\GeV}{M_{\rm SUSY}}^2\mytag. 
\end{align}
\restore
for the chargino and gluino one loop contribution to $d_d$
respectively.
In this and in the following estimates, we neglect
all the numerical coefficients of order one,
we choose $M_0\approx 10^{16}\GeV$ and
we use central values for the various known parameters, in
particular, $\J{CP}=2\cdot 10^{-5}$.
Moreover `$m_{\rm SUSY}$' stands, in each case, for the
particular combination of soft parameters written
in the analytical approximation.
Because of $d_u/d_d\circa{<}
(\tan\beta/55)^3$ and of $x/y\approx 7$ in eq.~\eqref{dn2}, the
corresponding contributions to $d_u$ are less interesting and totally
negligible for moderate $\tan\beta$.
At a closer inspection, these estimates turn out to be correct
within an order of magnitude.

As it is apparent from eq.~\eqref{dnt}, $d_n$ cannot be much larger
than $10^{-33}\ecm$ for $\tan\beta=10$ and than $10^{-31}\ecm$ for
whatever value of $\tan\beta$.
We remark that, owing to the necessary presence of $\lambda^2_c$ in
all results for $d_d$, a numerical calculation of the $d$ quark EDM
requires the knowledge of the 
low energy parameters with a precision of about $1/10^6$
in order to see the reciprocal cancellation of the terms not
proportional 
to $\lambda^2_c$.

\medskip

It is at this point interesting to understand why
the three-loop pure SM contributions to the neutron EDM, $d_n^{\rm
SM}\approx 10^{-32}\ecm$~\cite{dNreview}, are not suppressed relative
to the one-loop supersymmetric ones.

A large suppression of the pure SM contribution would seem plausible
because, in comparison with it, the supersymmetric one loop
contributions can be enhanced by
\begin{enumerate}
\item large logarithms, $\log M_0^2/M^2_Z$, that
multiplicate the loop
factors $(4\pi)^{-2}$ in the RGE inserctions involved by the one loop
supersymmetric diagrams;
\item $\tan\beta$ factors, that can enhance the $\lad$ and
$\lae$ couplings;  
\item a particularly small effective
combination of SUSY parameters, `$m_{\rm SUSY}$'. 
\end{enumerate}

On the other hand, the SM contributions are enhanced in different
ways. In fact, in the pure SM graphs,
\begin{enumerate}
\item[i.] the elicity flip can occur on an external leg, giving a factor
equal to the ``constituent quark mass'' $m_q^{\rm const}\sim m_n/3$,
equal for $q=\{u,d,s\}$
(so that $d_d^{\rm SM}\approx d_s^{\rm SM}$);
\item[ii.] the Yukawa couplings are expressed in terms of quark masses
evaluated at low energy and therefore enhanced by QCD renormalization
factors, as the strong coupling;  
\item[iii.] unlike the RGE-induced contributions, that must depend in a
polinomial way on the Yukawa couplings, the infrared structure
due to subtractions of quark propagators
in the SM graphs, give rise to
a mild logarithmic dependence.
\end{enumerate}
Point iii.\ is particularly important in the case of
quantities neutral under the quark ${\rm U}(3)^3$ flavour group
(like the electron EDM, or the phase of the $B$-term,
or the strong CP angle, or the 3-gluon operator~\cite{WeinbergGGG})
that for this reason receive purely supersymmetric contributions
much smaller than SM ones.
Due to the absolute lack of experimental interest,
we avoid discussing such issues in any detail.

Other than one loop supersymmetric and three loop pure SM diagrams, it
is also possible to have higher loop contributions, not considered
before, having the two kind of enhancement. An example can be obtained
from the chargino one loop diagram by adding by adding a QCD loop,
that allows elicity flip on the external lines and gives a
contribution to $d_d$ comparable with the previously computed one.
Using charged Higgs exchange, it is also possible to obtain an
interesting three loop contribution to $d_n$ given by 
\begin{equation}
d_u^{H^\pm}\approx e \frac{\alpha_3}{(4\pi)^5}\frac{m_n
m_t^2}{m_{H_\pm}^4} \lambda_b^2 \lambda_s^2 J_{\rm CP}
\approx10^{-32}\:\ecm\:\fracwithdelims(){\tan\beta}{55}^4
\fracwithdelims(){200\GeV}{m_{H_\pm}}^4.
\label{conc}
\end{equation}
Fine tuning considerations suggest that, in the large $\tan\beta$
region, the charged Higgs are lighter than the squarks~\cite{LargeTan}.

To conclude this section, let us note that the universality scale
$M_0$ can be much smaller than the unification scale if the
supersymmetry breaking is transmitted at relatively low energy by
gauge interactions.
In this case all RGE-induced contributions to $d_n$ are smaller
because of their strong dependence on $t_Z$, whereas the higher loop
ones, and in particular \eqref{conc}, are not.

\section{EDMs in unified theories}
\label{sec:unified}
In this section, we suppose that the gauge group is unified at the
universality scale $M_0$ that we will identify with the reduced Planck
mass.
In this case, EDMs can be much larger than in the MSSM.
As seen in the previous section, the smallness of EDMs in the MSSM is
due to the light Yukawa couplings, that have to be present to prevent
the removal of all phases from the lagrangian, phases that at the Planck
scale reside only in the Yukawa matrices. In fact, with regard for
example to the
quark $d$, if $\lambda_u=\lambda_c=0$ CP violation can
be removed with independent redefinitions of the right down quark and
of the left doublet.
If the gauge group is unified, at the Planck scale the phases still
reside 
only in the Yukawa matrices, but in this case MSSM multiplets
belonging to the same representation of the unified group can no
longer be rotated independently. Hence, depending on the gauge group
and on the quark in consideration, CP violation can persist even if
some or all of light Yukawa couplings are vanishing, making not
necessary their suppressing presence in the EDM expressions.

Let us consider first  minimal SU(5) unification. One generation is
composed by two multiplets, a five-plet $\overline{F}=(d^c,L)$ and a
ten-plet 
$T=(u^c,e^c,Q)$, and the Higgs fields belong to two five-plets,
$\bar{H}$, that transforms like $\overline{F}$, and $H$, that
transforms in the conjugate way.
In terms of these fields, the Yukawa interactions are 
\begin{equation}\label{yuk}
\frac{1}{4}\:T_i\lambda_{u_i} T_i\, H,\qquad
\sqrt{2}\: \overline{F}_i\lambda_{d_i}V^{\dagger}_{ij} T_j \,\bar{H}
\end{equation}
in a basis in which $\lau$ is diagonal. Below the unification scale,
$\lau$ and $\lad$ are the same of the previous section.

Five of the six phases of the CKM matrix can be removed in the MSSM
through independent redefinitions of the $Q_i$, $u^c_i$,
$d^c_i$. Here, it 
is easy to show that only three phases can be removed by independent
transformations of $T_i$, $\overline{F}_i$, so there are two more
physical phases 
compared to the MSSM (and the SM). Whereas two phases
decouple below the unification scale in non supersymmetric models, in
supersymmetric models they leave their effects in soft breaking
parameters, that loose their universality and reality, before
decoupling could make them harmless.

Is the effect of these two further phases on the EDMs suppressed by
light Yukawa couplings in this model?
The answer is no for the quark $u$ and yes, but in a different way
compared to the MSSM, for down quarks and charged leptons.
In order to show that, let us suppose first that the two light $\lau$
eigenvalues, $\lambda_u$ and $\lambda_c$, are vanishing. In this case,
all phases in the Yukawa matrices become unphysical. In fact, the
first of the two interactions~\eqref{yuk} becomes $T_3\lambda_t T_3
H/4$ and it is invariant for transformations in the $T_1$-$T_2$
sector.
Moreover, as it is well known, it is possible to write the most
general unitary matrix $V^\dagger$ as 
\begin {equation}
V^\dagger=
\text{diag}(e^{i\alpha},e^{i\beta},e^{i\gamma})
\left(\begin{array}{c|c}
\boldsymbol{R}_1&\boldsymbol{0}\\\hline\boldsymbol{0}&1
\end{array}\right)
\left(\begin{array}{c|c}
e^{i\delta}&\boldsymbol{0}\\\hline\boldsymbol{0}&\boldsymbol{R}_2
\end{array}\right)
\left(\begin{array}{c|c}
\boldsymbol{R}_3&\boldsymbol{0}\\\hline\boldsymbol{0}&1
\end{array}\right)
\text{diag}(e^{i\alpha'},e^{i\beta'},1),
\end{equation}
where $\boldsymbol{R}_1,\boldsymbol{R}_2,\boldsymbol{R}_3$ 
are orthogonal $2\times2$ matrices. Then
we can reabsorb
$e^{i\alpha},e^{i\beta},e^{i\gamma},e^{i\alpha'},e^{i\beta'}$ by a
redefinition of
$\overline{F}_1,\overline{F}_2,\overline{F}_3,T_1,T_2$, reach the
phase $e^{i\delta}$ 
by defining $(T'_1,T'_2)^T= \boldsymbol{R}_3(T_1,T_2)^T$ and reabsorb also
$e^{i\delta}$ by a redefinition of $T'_1$, without affecting the
diagonal interaction.
Since, as seen in the previous section, the limit
$\lambda_u=\lambda_c=0$ is meaningful for the down quarks and for the
charged leptons (and for the top), this means that their EDMs are
suppressed 
by light up Yukawa couplings in some way.

Let us suppose now that the two light $\lad$ eigenvalues, $\lambda_d$
and $\lambda_s$, are vanishing. In this case, two phases remain
physical. In fact, the interactions~\eqref{yuk} become 
\[
\frac{1}{4}\: T_i\lambda_{u_i}T_i\, H,\qquad
\sqrt{2}\:\overline{F}_3\lambda_b V^{\dagger}_{3j}T_j\,\bar{H}. 
\]
With a redefinition of $\overline{F}_3$, one of the three phases of
$V^\dagger_{31},V^\dagger_{32},V^\dagger_{33}$ can be reabsorbed, but
not the remaining two because, if they were reabsorbed by a
redefinition of some $T_i$, they would reappear in the diagonal
interaction. The two remaining phases are just those ones related with
unification. Since the limit $\lambda_d=\lambda_s=0$ is meaningful for
up quarks (and the bottom and the tau), their EDMs are not
necessarily suppressed by light Yukawa couplings.

The dependence of EDMs on Yukawa couplings can be obtained as in the
MSSM. Changing basis in the flavour space for the SU(5)
supermultiplets $\hat{T}$, $\hat{\overline{F}}$ corresponds to make a
$\text{U(3)}_T\times\text{U(3)}_{\overline{F}}$ transformation on the
flavour component of the fields, relative to which the Yukawa couplings
transform in this way:
\begin{equation}
\label{uUT}
\lau\rightarrow {U_T}^T\lau U_T,\qquad \lad\rightarrow
{U_{\overline{F}}}^T\lad U_T.  
\end{equation}

As before, the down quark EDMs are given by $d_{d_i}=\im\bigl[D_{d^R_i
d^L_i}\bigr]$, where the matrix $D$ transforms as $\lad$.
Unlike the case of the MSSM, this does not mean that
$D=\lad f( 
\laud\lau, 
\ladd\lad)$.
In this case we have rather 
\begin{equation}\label{SU(5)}
D=\lad f(
(\ladd\lad),
(\ladd\lad)^{*},
\lau,\laun^{*}), 
\end{equation}
where $f$ depends on the arguments in such a way that $f\rightarrow
U_T^{\dagger}f U^{}_T$ for a transformation~\eqref{uUT}.
From eq.~\eqref{SU(5)} it follows that $d_d$ is proportional to
$\lambda_d$ as in the MSSM.
Moreover, $f$ must depend on down Yukawa couplings through their
squares but it can depend on individual up Yukawa couplings also
through 
not squared couplings, provided that the total number of $\lau$ is
even. This can be seen by noting that the particular flavour
transformation $T\rightarrow
iT$, $\overline{F}\rightarrow \overline{F}$ leaves $\ladd\lad$ and $f$
unchanged but it changes sign to 
$\lau$. Therefore, unlike the MSSM, a $\lambda_c$ suppression can be
enough for the quark $d$. Indeed, the first non vanishing
contribution to $d_d$ is proportional to 
\begin{equation}\label{pippo}
\im \bigl[\bigl(\lad\laud
(\ladd\lad)^{*}
\laun^3\bigl)_{d_R d_L}\bigr]=
\lambda^{}_d\lambda_t^3\lambda_b^2\lambda^{}_c
\im\bigl[V^{\dagger}_{dc}\overline{V}^{}_{cb}V^T_{bt}V^{}_{td}\bigr].
\end{equation}
The EDMs of the quark $s$ and of the electron are suppressed in an
analogous way.

The up quark EDMs behave in a different way. They are given by
$d_{u_i}=\im\bigl[U_{u^R_i u^L_i}\bigr]$ where the matrix $U$
transforms as $\lau$ under the transformation~\eqref{uUT}. Unlike
the MSSM, $U$ is not necessarily proportional to $\lau$ on the left,
but can also be proportional to $\ladn^T$, that transforms in the same
way on the left. Therefore, $d_u$ is not necessarily proportional to
$\lau$, giving rise to a possible enhancement.
In any case, $d_u$ must contain an odd number of up Yukawa matrices,
since the transformation $T\rightarrow iT$, $\overline{F}\rightarrow
\overline{F}$ leaves 
$\lad$ unchanged but it changes sign to $\lau$ and $U$. The first non
vanishing contribution to $d_u$ is proportional to 
\begin{equation}
\label{ddu}
\im\bigl[(\ladd\lad)^{*}\lau(\ladd\lad)\bigr]=
\lambda^{}_t\lambda_b^4
\im\bigl[\overline{V}^{}_{ub}V^T_{bt}V_{tb}V^{\dagger}_{bu}\bigr]
\end{equation}
and it exhibits the described features.
In eqs.~(\ref{pippo},\ref{ddu}) we omitted the renormalization factors for
the $V_{\text{CKM}}$ matrix elements.

\begin{figure}
\begin{center}
\epsfig{file=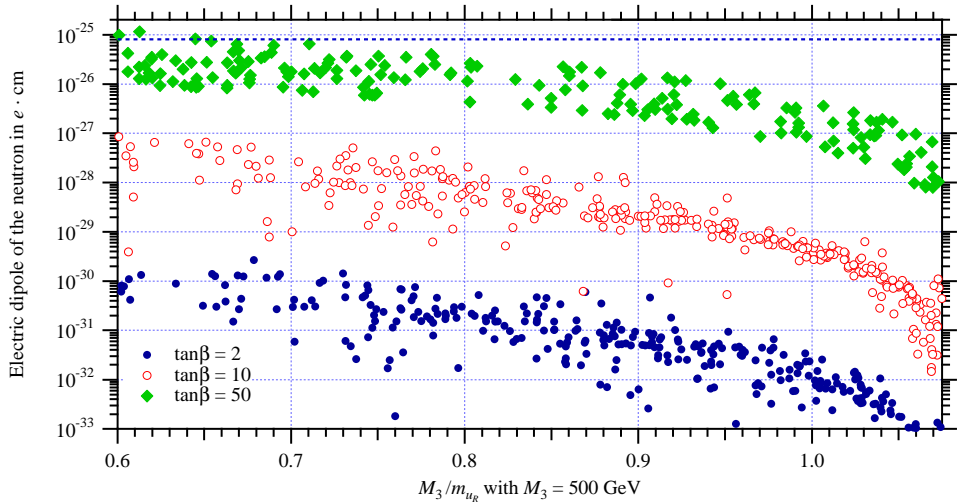,width=0.9\textwidth}
\end{center}
\caption{\em The contribution to the neutron EDM generated by minimal
SU(5) for $M_\glum=500\GeV$, $\tan\beta=2,10,50$ and
$|\im[V^2_{tb}\overline{V}^2_{ub}]|=10^{-5}$, as a function of the
ratio between gaugino and scalar masses and for random samples of
acceptable sparticle spectra.}
\label{fig:du}
\end{figure}

In SU(5) we have therefore $d_u\gg d_d,d_s,d_e$. Only the gluino
diagram is able to give rise to the $\lambda_t$ enhancement. In fact,
the chargino diagram is explicitely proportional to the $\lambda_u$
that 
appears in one of its vertices and it is negligible in this model.
Let us concentrate then on the gluino contribution to the quark $u$,
$d^{\glu}_u$. The flavour
structure~\eqref{ddu} is generated by
the corrections to the mass matrices.
At the unification scale, these corrections are proportional to
combinations of Yukawa matrices 
transforming as the mass matrices themselves relative to the
$\text{U(3)}_T\times\text{U(3)}_{\overline{F}}$ group. Hence the
corrections to the 
ten-plet mass matrix, and so those to $m^2_Q$, ${m^2}_{u_R}^{\,*}$,
${m^2}_{e_R}^{\,*}$, are proportional to $\boldsymbol{1}$, $\laud\lau$,
$\ladd\lad$, etc., whereas the
corrections to the five-plet mass matrix, and so those to $m^2_L$ and
${m^2}_{d_R}^*$, are proportional to $\boldsymbol{1}$,
$(\lad\ladd)^*$, etc.. All these corrections are also proportional to
$t_{\mathrm{G}}\equiv(4\pi)^{-2}\log(\M{P}^2/\M{G}^2)\simeq
0.06$. Some of them are characteristic of unification, whereas other
ones are produced also below the unification scale and hence at the
Fermi scale they are proportional to
$t_Z=(4\pi)^{-2}\log(\M{P}^2/M^2_Z)\simeq 0.5$. Including also some
approximate numerical factors, the corrections $\Delta m^2$ at the Fermi
scale are
\begin{alignat}{2}
\label{mQZ}
&\Delta m^2_{\tilde{\ped{Q}}}&\mbox{}\propto\mbox{} &\boldsymbol{1}
-3 t_Z\laud\lau
-3 t_Z\ladd\lad \mytag\\
&\Delta m^2_{\tilde{\ped{u}}_R} &\mbox{}\propto\mbox{} &\boldsymbol{1}
-6 t_Z\lau\laud 
-6 t_{\mathrm{G}}(\ladd\lad)^*
\mytag\\
&\Delta m^2_{\tilde{\ped{d}}_R} &\mbox{}\propto\mbox{} &\boldsymbol{1}
-6 t_Z\lad\ladd. 
\mytag
\end{alignat}
\restore
These corrections are able to generate the flavour dependence of
eq.~\eqref{ddu}, so that we can estimate $d_n$ as
\begin{equation}\label{dne}
\begin{split}
|d_n|&\approx
e\frac{\alpha_3}{4\pi}\lambda_b^4 
\im\bigl[\overline{V}^{}_{ub}V^T_{bt}V^{}_{tb}V^{\dagger}_{bu}\bigr]
t_Zt_{\mathrm{G}}\frac{m_tA_u}{m^2_{\tilde{u}}}
\frac{M_\glum}{\max(m^2_{\tilde{u}}, M^2_\glum)}
\\ 
&\approx 10^{-(25\div26)}\fracwithdelims(){\tan\beta}{50}^4
\fracwithdelims(){|\im[V^2_{tb}\overline{V}^2_{ub}]|}{10^{-5}}
\fracwithdelims(){500\GeV}{M_{\text{SUSY}}}^2\ecm,
\end{split}
\end{equation}
where $M_{\text{SUSY}}$ in the second line is the combination of
supersymmetric parameters appearing in the first line
and it can be even larger than $500\GeV$ because of the likely
lightness of $A$ in the large $\tan\beta$ regime.
Since the prediction for $d_u$ is interesting enough, we use the
expressions~\eqref{expr} to do an exact computation of $d_n$ that we show
in figure~\ref{fig:du} for
$M_\glum=500\GeV$, for three values of $\tan\beta$, 2, 10 and 50, and for
$|\im[V^2_{tb}\overline{V}^2_{ub}]|=10^{-5}$. 
$d_u$ is plotted as a
function of $M_\glum/m^2_{\tilde{u}_R}$ in order to exhibit the
so-called ``gluino focussing'' effect~\cite{barbieri:95a}.

From eq.~\eqref{dne} and Fig.~\ref{fig:du} and from the experimental
limit $d_n<0.8\cdot 10^{-25}\ecm$~\cite{altarev:92a} it is apparent
that the $u$ quark EDM can be very large in superunified theories with
large $\tan\beta$. 

Let us consider now shortly the gauge group SO(10), whose sixteen
dimensional representation ``$16$'' unifies all quarks and leptons of
one 
generation, included a right-handed neutrino. In the minimal model the
two light Higgs doublets belong to two different ten dimensional
SO(10) multiplets, $10_\ped{u}$ and $10_\ped{d}$, and the Yukawa
interactions are 
\begin{equation}
\label{uiS}
16^T\:\lau\: 16\: 10_{\ped{u}}, \qquad 
16^T\:\lad\: 16\: 10_{\ped{d}}.
\end{equation}
In this model the situation is different from the SU(5) case, in which
the five-plets and ten-plets could be rotated independently.
All six phases of $V$ are physically significant. Moreover, it is easy
to see that in this case it is not possible to remove the CKM phases
neither when $\lambda_d=\lambda_s=0$ (like in SU(5)) nor when
$\lambda_u=\lambda_c=0$ (unlike in SU(5)). Therefore neither $d_u$ nor
$d_d$ are suppressed by light Yukawa couplings.

The $u$ quark EDM is generated in the same way it is generated in
SU(5) and its estimate is identical.
On the other hand, whereas in SU(5) only $m^2_{u_R}$ (and not
$m^2_{d_R}$) has non MSSM corrections, in SO(10) the situation is
symmetric under $u$-$d$ exchange, so that $m^2_{d_R}$ has
corrections depending on $\lad$ as the $m^2_{u_R}$ ones depend on
$\lau$. Moreover, also an 
approximate expression for $d_d$ can be obtained from that one for
$d_u$ exchanging $u$ and $d$. 
$d_d$ has therefore a linear dependence on $\tan\beta$. Moreover, when
$\tan\beta$ is large and $\lambda_b\simeq\lambda_t$ the main
differences between $d_u$ and $d_d$ are due to the different
combination of CKM angles and phases,
$\im\bigl[\overline{V}^2_{td}V^2_{tb}\bigr]$ instead of
$\im\bigl[\overline{V}^2_{tb}V^2_{ub}\bigr]$, and to the different RL
flip, that is provided by $\mu$ and not $A_u$. With similar phases,
the CKM angles could favour $d_d$. Nevertheless, also in SO(10) with
large $\tan\beta$ the $u$ quark contribution cannot be forgotten.

It is also interesting to estimate the corrections to the quark masses
that, in SU(5) and in SO(10), contribute to the strong CP
violating angle $\theta$
in models without an axion.
These contributions are 
\begin{equation}
\delta\theta^u_{\text{QCD}}\approx
\frac{\alpha_3}{4\pi}\frac{m_t}{m_u} \lambda_b^4
\im\bigl[\overline{V}^2_{ub}V^2_{tb}\bigr]t_Zt_{\mathrm{G}}
\approx 10^{-11}\tan^4\!\beta
\fracwithdelims(){|\im[\overline{V}^2_{ub}V^2_{tb}]|}{10^{-5}} 
\end{equation}
both in SU(5) and SO(10) and
\begin{equation}
\delta\theta^d_{\text{QCD}}\approx
\frac{\alpha_3}{4\pi}\frac{m_b}{m_d} \lambda_t^4
\im\bigl[\overline{V}^2_{td}V^2_{tb}\bigr]t_Zt_{\mathrm{G}}
\frac{\mu M_\glum}{m^2_{\tilde{u}}} 
\approx 10^{-5}\tan\beta
\fracwithdelims(){|\im[\overline{V}^2_{td}V^2_{tb}]|}{10^{-4}}
\end{equation}
in SO(10).
The bound $\theta_{\text{QCD}}\leq 10^{-9}$ shows that an axion is
necessary in SO(10) and also in SU(5) if $\tan\beta$ is large. 
It is interesting that
the large amount of CP violation left by SO(10) unification
in the soft terms
furnishes an experimental possibility
to see the `invisible' axion~\cite{axion}.

\section{Conclusions}
\label{sec:con}
We have studied EDMs produced by the misalignment  of two Yukawa
matrices in models with universality of soft breaking terms at an high
energy scale from the point of view of transformation properties under
flavour transformation commuting with the gauge group. In this way it
is simple to get the dependence of EDMs from Yukawa couplings,
summarized in table~\ref{tab:summary}.
On this basis, estimates are possible and effective
for small and large effects. 
\begin{table}
\renewcommand{\arraystretch}{1.3}
\begin{center}
\[
\begin{array}{|c|ccc|}
\hline 
& \text{MSSM} & \text{minimal SU(5)} & \text{\rm minimal SO(10)}\\
\hline 
\,d_u &\lambda_u \lambda_b^4 \lambda_t^2 \lambda_s^2 \J{CP}
    &\lambda_t \lambda_b^4 \im[\overline{V}^2_{ub}V^2_{tb}]
    &\lambda_t \lambda_b^4 \im[\overline{V}^2_{ub}V^2_{tb}]\\
\,d_d &\lambda_d \lambda_t^4 \lambda_b^2 \lambda_c^2 \J{CP} 
    &\lambda_d \lambda_t^3 \lambda_b^2 \lambda_c
     \im[\overline{V}_{cd}\overline{V}_{cb}V_{tb}V_{td}]
    &\lambda_b \lambda_t^4 
     \im[\overline{V}^2_{td}V^2_{tb}]\\
\,d_e &\lambda_e \lambda_t^4 \lambda_b^4\lambda_s^2\lambda_c^2 \J{CP}
    &\lambda_e \lambda_t^3 \lambda_b^2 \lambda_c 
     \im[\overline{V}_{cd}\overline{V}_{cb}V_{tb}V_{td}] 
    &\lambda_\tau \lambda_t^4 
     \im[\overline{V}^2_{td}V^2_{tb}]\\
\hline 
\end{array}
\]
\end{center}
\caption{\em Flavour factors suppressing the purely
supersymmetric contributions to light fermion EDMs
in different minimal models.}
\label{tab:summary}
\end{table}

In the case of MSSM, owing to light Yukawa coupling suppression, the
effects given by one loop diagrams are largely below present
experimental limits (see eq.~\eqref{dnt}). Therefore, the detection of
a non vanishing EDM 
for the neutron in foreseeable experiments would be an inequivocable signal
of physics beyond the MSSM. Some interesting contribution to $d_n$ can
arise also from more loop diagrams.

In the case of minimal supersymmetric SU(5) and large $\tan\beta$, on
the contrary, the effects can be close to
experimental limits with regards to neutron EDM because of the
contribution of $u$ quark, proportional to $\tan^4\!\beta$. For this
interesting case the results of an exact computation are given by
Fig~\ref{fig:du}.
The smallness of $d$ quark and electron 
EDMs compared with the $u$ quark one is also characteristic of
SU(5). Therefore SU(5) can be distinguished from the MSSM because it
can give rise to a measurable EDM for the neutron. Moreover, SU(5)
effects can be 
distinguished from the SO(10) ones and from effects coming from
universal 
intrinsic phases in $A$ and $B$ terms~\cite{barbieri:96a} because of
the smallness of the electron EDM. On the other hand, at least for the
EDMs a similar pattern
can be given by strong CP violation. 

In the case of minimal supersymmetric SO(10) it is well known that the
$d$ quark and electron EDMs can be close to the experimental
limits. If $\tan\beta$ is large, the effect is enhanced by a factor
$\mu\tan\beta/A_d$. Also in this case, as in
SU(5), the $u$ quark gives an important contribution that must be
taken in account. 

\section*{Acknowledgements}
We thank R. Barbieri for conversations and help. 

\appendix
\renewcommand{\theequation}{A\arabic{equation}}
\setcounter{equation}{0}
\section*{Appendix: general formulas for one loop EDMs}\label{sec:neu}

It is possible to express the neutron EDM $d_n$ in terms of the
contribution of the quark $q$ to the neutron spin 
$(\Delta q)_n$ and of 
its EDM $d_q$~\cite{ellis:96a}:
\begin{equation}
\label{dn1}
d_n = \zeta (\Delta d)_n d_d +\zeta (\Delta u)_n d_u + \zeta (\Delta
s)_n d_s, 
\end{equation}
where $(\Delta d)_n = 0.82\pm 0.03$, $(\Delta u)_n = -0.44\pm 0.03$ e
$(\Delta s)_n = -0.11\pm 0.03$ \cite{ellis:96a} and $\zeta$ is a
renormalization factor, $\zeta\simeq 1.6$. 

Concerning the quark EDMs, 
let us consider a supersymmetric extension of Standard Model at the
Fermi scale, with minimal field contents, imposed R-parity and generic
soft terms a part from a reality hypothesis on $B$ (defined as the
coefficient that appears in the soft breaking term $B\mu H_uH_d$,
where $\mu$ is the coefficient of the corresponding interaction in the
superpotential) from which it
follows that the chargino's and neutralino's mixing matrices can be
taken real. Such hypothesis is
verified with very good  precision in the models that we will
consider. Then the full
expressions for the one loop EDMs of up and down quarks are 
\begin{align}\label{expr}
 d_{u_i} = &
\mbox{}-\frac{e}{(4\pi)^2}
H^-_{n \tilde{h}^-_\ped{d}} \frac{1}{M_{\chi_n}}
H^+_{n\tilde{h}^+_\ped{u}} \im\Big{[}(\ladd f^{\chi}_u
(\frac{M^2_D}{M^2_{\chi_n}})_{RL} \laud)_{u^L_i u^R_i}\Big{]}\notag\\
\mbox{}&+\frac{8}{3} e \frac{\alpha_3}{4\pi}
\frac{1}{M_\glum} \im\Big{[}f^\glu_\ped{u} 
(\frac{M^2_U}{M^2_\glum})_{u^L_i u^R_i}\Big{]}\tag{\ref{expr}a} \\
\mbox{}&+ \frac{y(\ped{u}_R)}{3\cos^2\!\theta_W} e
\frac{\alpha}{4\pi} 
H_{n\tilde{B}}\frac{1}{M_{N_n}}
(\frac{t_3(\ped{u}_L)}{y(\ped{u}_L)}\cot\theta_W H_{n\tilde{W}_3} +
H_{n\tilde{B}})
\im\Big{[}f^{N}_\ped{u} 
(\frac{M^2_U}{M^2_{N_n}})_{u^L_i u^R_i}\Big{]}\notag\\ 
\mbox{}&+\lambda_{u_i}^2 \frac{e}{(4\pi)^2} 
H_{n \tilde{h}_\ped{u}^0} \frac{1}{M_{N_n}} 
H_{n \tilde{h}_\ped{u}^0}
\im\Big{[}f^{N}_\ped{u}(\frac{M^2_U}{M^2_{N_n}})_{u^L_i u^R_i}\Big{]}
\notag\\
 d_{d_i} = &
\mbox{}-\frac{e}{(4\pi)^2}
H^+_{n \tilde{h}^+_{\mathrm{u}}} \frac{1}{M_{\chi_n}}
H^-_{n\tilde{h}^-_{\mathrm{d}}}
\im\Big{[}(\laud
f^{\chi}_\ped{d}(\frac{M^2_U}{M^2_{\chi_n}})_{RL} 
\ladd)_{d^L_i d^R_i}\Big{]}\notag\\
\mbox{}&+\frac{8}{3} e \frac{\alpha_3}{4\pi}
\frac{1}{M_\glum} \im\Big{[}f^\glu_\ped{d}
(\frac{M^2_D}{M^2_\glum})_{d^L_i d^R_i}\Big{]}\tag{\ref{expr}b}\\
\mbox{}&+ \frac{y(\ped{d}_R)}{3\cos^2\!\theta_W} e \frac{\alpha}{4\pi} 
H_{n\tilde{B}}\frac{1}{M_{N_n}}
(\frac{t_3(\ped{d}_L)}{y(\ped{d}_L)}\cot\theta_W H_{n\tilde{W}_3}
+H_{n\tilde{B}})
\im\Big{[}f^{N}_\ped{d}(\frac{M^2_D}{M^2_{N_n}})_{d^L_i d^R_i}\Big{]}
\notag\\  
\mbox{} &+\lambda_{d_i}^2\frac{e}{(4\pi)^2}
H_{n \tilde{h}_\ped{d}^0}\frac{1}{M_{N_n}} 
H_{n \tilde{h}_\ped{d}^0}
\im\Big{[}f^{N}_\ped{d}
(\frac{M^2_D}{M^2_{N_n}})_{d^L_i d^R_i}\Big{]},\notag
\stepcounter{equation}
\end{align}
where $t_3(a)$ and
$y(a)$ are respectively 
the third component of weak
isospin and the hypercharge of the particle $a$ (normalized in such a
way that $q=t_3+y$). $\glu$ is the gluino
and $M_\glum$ its mass,
$N_n$, $n=1\ldots 4$ 
are the neutralinos, $\chi_n^+$, $\chi_n^-$, $n=1\ldots 2$ the
charginos and $H$, $H^+$, $H^-$ their mixing matrices. $M^2_U$,
$M^2_D$ are the $6\times 6$
squarks mass matrices in the mass eigenstate basis of corresponding
quarks 
\begin{equation}\label{masses:eq}
\begin{aligned}
M^2_U &= \begin{pmatrix}
{\mybf{m}^2_{\ped{\tilde{Q}}}}^{\!(\ped{u})} + 
\mybf{M}^2_{\ped{u}} + D_{\tilde{\ped{u}}_L} \boldsymbol{1} & 
-(\mybf{A}_{\ped{u}}^{\dagger}+\mu\cot\beta \,\boldsymbol{1}) 
\mybf{M}_{\ped{u}} \\
-\mybf{M}_{\ped{u}} (\mybf{A}_{\ped{u}}+\mu\cot\beta \,\boldsymbol{1}) &
\mybf{m}^2_{\tilde{\ped{u}}_R} + \mybf{M}^2_{\ped{u}} +
D_{\tilde{\ped{u}}_R} \boldsymbol{1} 
\end{pmatrix}\\
M^2_D &= \begin{pmatrix}
{\mybf{m}^2_{\ped{\tilde{Q}}}}^{\!(\ped{d})} + 
\mybf{M}^2_{\ped{d}} + D_{\tilde{\ped{d}}_L} \boldsymbol{1} & 
-(\mybf{A}_{\ped{d}}^{\dagger}+\mu\tan\beta \,\boldsymbol{1}) 
\mybf{M}_{\ped{d}} \\
-\mybf{M}_{\ped{d}} (\mybf{A}_{\ped{d}}+\mu\tan\beta \,\boldsymbol{1}) &
\mybf{m}^2_{\tilde{\ped{d}}_R} + \mybf{M}^2_{\ped{d}} +
  D_{\tilde{\ped{d}}_R} \boldsymbol{1} 
\end{pmatrix}.
\end{aligned}
\end{equation}
In the right sides of~\eqref{expr} we used a matrix 
notation. The $f$s are the appropriate loop functions, namely
linear combinations $q_s g_2 + q_f h_2$, where $q_s$ and $q_f$ are the
electric charges of the particles running respectively in the scalar and
fermion line of the corresponding diagram (in unity of $e$), and 
\begin{equation}\label{loop}
      g_2(r) = \frac{1}{2(r-1)^3}[r^2-2r\log r -1],\qquad
      h_2(r) = \frac{1}{2(r-1)^3}[-2r^2\log r +3 r^2-4r+1].
\end{equation} 

In eqs.~\eqref{expr} the first two contributions to EDMs are due to
charged higgsinos and 
gluinos exchange and they are the dominant ones. The third ones take
account of bino and neutral wino exchange and they are less important
than the corresponding gluino exchange, while the last ones come from
neutral higgsino exchanges and they are completely negligible.
All the contributions in eqs.~\eqref{expr} come from diagrams with none
or two elicity flips on the vertices. The one loop diagrams in which
the elicity flip occurs on an external leg are always real. As such
they do not give any contribution. The diagrams with one elicity flip
on the vertices would contribute only if the $B$ term were complex. In
this case in eqs.~\eqref{expr} also the mixing matrices $H$ would be
complex and they would appear within the imaginary parts.

The arguments of imaginary parts contain both the model dependence and
the difficulty of the calculation of EDMs.


\end{document}